\documentclass[a4paper,10pt]{article}

\usepackage{fullpage,amsmath,amssymb,amsfonts,epsfig, subfigure, float, setspace}
\renewcommand\thefootnote{\fnsymbol{footnote}}

\psfigdriver{dvips}

\begin{document}
\title{Diversity as a product of interspecial interactions}

\author{Daniel Lawson\footnotemark[1], Henrik Jeldtoft Jensen\footnotemark[1] \footnotemark[2] and Kunihiko Kaneko\footnotemark[3]}
\date{9.05.05}



\maketitle

\begin{abstract}
We demonstrate diversification rather than optimisation for highly interacting organisms in a well mixed biological system by means of a simple model and reference to experiment, and find the cause to be the complex network of interactions formed, allowing species less well adapted to an environment to flourish by co-interaction over the `best' species.  This diversification can be considered as the construction of many co-evolutionary niches by the network of interactions between species.  Evidence for this comes from work with the bacteria \emph{Escherichia coli}, which may coexist with their own mutants under certain conditions.  Diversification only occurs above a certain threshold interaction strength, below which competitive exclusion occurs.
\end{abstract}

\section{Introduction}

\fnsymbol{footnote}
\footnotetext[1]{Department of Mathematics, Imperial College London,  South Kensington campus,  London SW7 2AZ, U.K.}
\footnotetext[3]{Department of Basic Science, University of Tokyo and ERATO Complex Systems Biology, JST, Komaba, Meguro-ku, Tokyo 153-8902, Japan}
\footnotetext[2]{Author for correspondence (h.jensen@imperial.ac.uk)}
\renewcommand\thefootnote{\arabic{footnote}}

Understanding how diversity arises through evolution and is sustained in an ecosystem is an important issue.  For some ecosystems, as seen in tropical rain forests, the diversity is rather high; it is quite low in other ecosystems. One of the key questions therein is whether interactions between organisms enhance or suppress diversity. If there is no explicit symbiotic interaction, it would be expected that the competition for a given resource leads to exclusion of many types, leading to the monodominance, i.e., the survival of the fittest, as determined by Gause's competitive exclusion principle \cite{Biology}.  In contrast, in a recent experiment by Kashiwagi et al. \cite{DiverseEcoli}, it was observed that diversification of several types of \emph{Escherichia coli} (or \emph{e. coli}) are facilitated under stronger interaction, for a well-mixed liquid culture with a given nutrition source.  In the experiment the sole nutrition source (of nitrogen) in the culture is glutamate, and through mutagenesis, evolution of a single gene was studied - the gene for glutamine synthetase production, which synthesises glutamine from glutamate.  Since the glutamine synthesis is necessary for the growth of the bacteria in this experiment, those with the higher activity of glutamine synthetase will result in faster growth of the bacteria.  Indeed, in a low population density condition, only the fittest, i.e., those with highest enzyme activity survives.  However, in a dense condition, multiple types including those with much lower enzyme activity coexist.  Kashiwagi et al. have also confirmed \cite{PrivCom} that the interaction is essential to coexistence of several types by cutting off a chemical in the medium that is supposed to be relevant to interaction, and showing the recovery of the survival of only the fittest.

Motivated by this experimental result, we show that the diversification is indeed facilitated by the interaction.  We do this by adopting a slightly modified version of the Tangled Nature (TaNa) model \cite{TaNaBasic, TaNaQuasi, TaNaTimeDep}. In addition to the standard, interspecial interaction in the TaNa model, we introduce differing values of self-interaction for each genotype. For comparison to the \emph{e.coli} experiment of Kashiwagi et al., this can be thought of as glutamine synthetase efficiency. A self-supporting, dominant genotype may coexist with, or be displaced by, a number of other genotypes that are less efficient competitors for the resource individually, provided that strong enough interactions are permitted. Diversity is maintained via the complex network of interactions.

The TaNa model is a simple individual based model, emphasising co-evolutionary aspects in ecology.  In particular, diversity occurs in the absence of space and without assumptions about the neutrality of species \cite{NeutralTheory, NTecologyNature, McKaneReview}, which although give very good matches with species abundance data \cite{SpecDiv} cannot answer broader questions about why such accuracy is achieved, as species are observed to interact strongly.  There are clearly aspects of evolution that require the specifics to be considered, and the key is finding an appropriate level of description.  The diversification observed in \emph{e.coli} is describable at the general interaction level of our model but not below, and although more specific models can be considered it is not necessary to do so for an understanding in general terms of the relationship between diversity and interaction.

\section{Definition of the Model}
\label{SecDefinition}

We now define the Tangled Nature model.  Individuals are represented as a vector $\mathbf{S}^{\alpha} = (S^\alpha_1,S^\alpha_2, ... , S^\alpha_L)$ in genotype space $\cal{S}$.  The $S^\alpha_i$ take the values $\pm 1$, and we use $L=20$ throughout.  Each $\mathbf{S}$ string represents an entire species with unique, uncorrelated interactions.  The small value of $L$ is necessary for computational reasons as all organisms exist \emph{in potentia} and have a designated interaction with all other species\footnote{When discussing the model, we refer to points in genotype space as species.  It is a matter of interpretation whether we consider genotype space to be `coarse-grained' (resulting in each being a different species - valid when $k$ and $\epsilon$ are `large' so that genotype differences affect reproduction probability greatly; see Equation \ref{HdefBasic} for definitions), or whether we consider genotype space to be a small sample of a much larger space (constant for all considered individuals), meaning genotypes are types of a base species (which would be valid when $k$ and $\epsilon$ are small, and so all genotypes have similar reproduction probabilities).  As we operate in neither extreme and reproduction is asexual, the distinction between species and type is difficult.}.

We refer to individuals by Greek letters $\alpha, \beta, ... = 1,2,..., N(t)$.  Points in genotype space are referred to as $\mathbf{S}^a, \mathbf{S}^b, ...$, and any number of individuals may belong to a point in genotype space $\mathbf{S}^a$.

In the original TaNa model, individuals $\alpha$ are chosen randomly and allowed to reproduce with probability $p_{off}$:

\begin{equation}
p_{off}(\mathbf{S}^\alpha,t) = \frac{\exp[H(\mathbf{S}^\alpha, t)]}{1+\exp[H(\mathbf{S}^\alpha, t)]} \in (0,1)
\end{equation}

They are then killed with probability $p_{kill}$, which is a constant parameter.  The difference between the original model and the one used here is the definition of the weight function $H(\mathbf{S}^\alpha, t)$.  The original version used was: 

\newfloat[H]{
\begin{equation}
\label{HdefBasic}
H_0(\mathbf{S}^\alpha,t) = \frac{k}{N(t)} \sum_{\mathbf{S} \in S} J(\mathbf{S}^\alpha, \mathbf{S}) n(\mathbf{S},t) - \mu N(t)
\end{equation}
}

Where $k$ ($=1/c$ from previous papers) is a control parameter, $N(t)$ is the total number of individuals at time $t$ and  $n(\mathbf{S},t)$ is the number of individuals with genotype $\mathbf{S}$ at that point.  The \emph{interaction matrix} $J(\mathbf{S}^\alpha, \mathbf{S})$ represents all possible couplings between all genotypes, each generated randomly in the range $(-1,1)$, being non-zero with probability $\Theta$.  Since the functional form of $J(\mathbf{S}^a,\mathbf{S}^b)$ does not affect the dynamics, provided that it is non-symmetric with mean $0$, we choose a form of the interaction matrix that speeds computation \cite{TaNaBasic}.  In the analysis sections, we will use shorthand versions: $J_{a b}$ as the interaction of an individual from species $b$ on an individual from species $a$, and $n_a$ as the number of individuals with genotype $a$.

The new model also allows a self-interaction term, giving the full weight function for an individual $\alpha$:

\newfloat[H] {
\begin{equation} 
\label{Hdef}
H(\mathbf{S}^\alpha,t) =H_0(\mathbf{S}^\alpha,t) + \frac{\epsilon}{N(t)} n(\mathbf{S}^\alpha,t) E(\mathbf{S}^\alpha  )
\end{equation}
}

Here, $\epsilon$ is a new control parameter and $E(\mathbf{S}^\alpha  )$ is the self-interaction of individual $\alpha$, generated uniformly random from the range $(0,1)$.  All members of the same species will have the same weight function and therefore the same offspring probability at a given time; i.e. if $\alpha$ was from species $a$ then $H(S^a) = H(S^\alpha)$.

To understand the meaning of the self-interaction term, we consider the weight function of a system with only one species $a$, $H(S^a) = \epsilon E(S^a) - \mu n(S^a)$ since $N=n(S^a)$.  If we assume that the system is in a steady state ($P_{off} = P_{kill}$), then $H(S^a)=H^* = -\ln(\frac{1}{p_{kill}}-1)$, which is constant.  Thus we find $n(S^a) = \frac{\epsilon E(S^a) +|H^*|}{\mu}$, meaning that $E(S^a)$ determines how successful species $S^a$ would be if alone in the system.  Hence consideration for negative $E(S^a)$ is not necessary.

Reproduction occurs asexually, and on a successful reproduction attempt two daughter organisms replace the parent, with each $\mathbf{S}^\alpha_i$ mutated (flipped from 1 to -1, or from -1 to 1) with probability $p_{mut}$.  Thus mutations are equivalent to moving to an adjacent corner of the L-dimensional hypercube in genotype space, as discussed in \cite{TaNaBasic}.

A time-step consists of choosing an individual\footnote{In previous versions a different individual was chosen for reproduction and killing actions.  Here we select only one individual and process it for reproduction and killing for code efficiency reasons - above the level of fluctuations the two methods are equivalent.} $\alpha$ randomly, and processing according to:

\begin{itemize}
\item
  $\alpha$ is allowed to reproduce with probability $p_{off}$.
\item
  $\alpha$ is killed with probability $p_{kill}$.  (if $\alpha$ reproduced, it is a daughter organism that is killed).
\end{itemize}

We define a generation as the amount of time for all individuals to have been killed, on average, once.  For a stable population size, this is also the time for all individuals to have reproduced once, on average.  The diversity is defined as the number of genotypes with occupancy greater than 20 to eliminate unsuccessful mutants from our count, and is called the wildtype diversity.

Unless otherwise stated, the parameters used will be: $\Theta=0.2$, $\mu=0.01$, $p_{mut}=0.015$, $\epsilon=2.0$ and $p_{kill}=0.1$; see \cite{TaNaBasic} for more details.  The initial conditions were determined by allowing the system to find a monodominant q-ESS by running the system for 5000 generations with all interaction disabled ($k=0$), thus the best competitor in the initial set was found.  Then the interaction was enabled by setting $k$ to the desired value.

\section{Results}

\subsection{Observed behaviour}

\begin{figure}
[htb]
\begin{center}
\epsfig{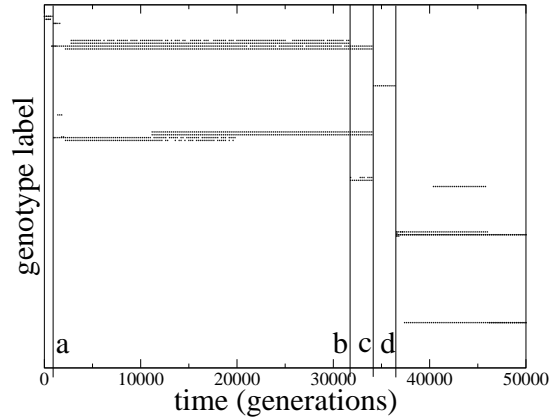}
\caption{\scriptsize{A sample run ($k = 10$) showing all genotypes with occupations greater than 20 as an unordered genotype label.  Times shown correspond to different cross-over types.  (a) is from the original monodominance to a diverse state, which 100 generations later becomes more diverse again.  (b) shows a cross-over from one diverse state to another, which at (c) becomes a new monodominant state.  Then at (d) the system returns to a new diverse state.} \label{FigExample}}
\end{center}
\end{figure}

As in the basic Tangled Nature model, the system experiences a number of quasi-Evolutionary Stable Strategies (or q-ESSs) during which a single genotype or set of genotypes is present with constant occupancy, ignoring fluctuations.  The q-ESS may end abruptly, leading to a disorder phase before a new q-ESS is found, which may or may not contain some of the same genotypes.  For the parameter ranges we study, the disorder phase lasts only tens of generations and so is instantaneous on an evolutionary timescale.  This behaviour is shown in Figure \ref{FigExample}, with some major events labelled.

\begin{figure}
[htb]
\begin{center}
\epsfig{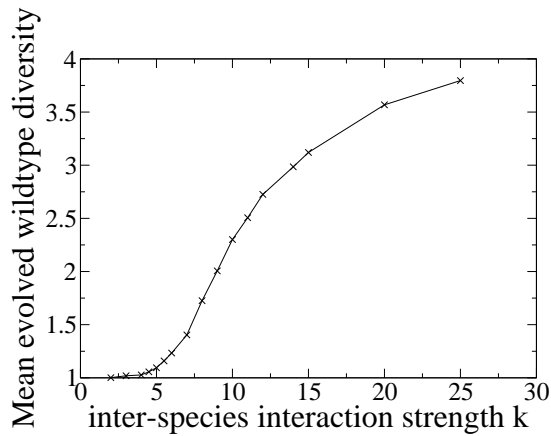}
\caption{\scriptsize{$k$ dependence of the average diversity of an evolved system, taken for $40000-50000$ generations and 500 runs per data point.} \label{FigKvsD}}
\end{center}
\end{figure}

We are interested in what happens as we change the interaction strength.  As shown in Figure \ref{FigKvsD}, around $k \approx 5.5$, a change from monodominance to the possibility of diverse systems occurs.  At low $k$ values, a large number of mutant species are examined by the system, but their numbers are suppressed by lack of positive interactions - if another species is successful then the original strain must become extinct.  At higher values of $k$ average diversity increases, with the possibility of both monodominance and diverse states at different times as shown in the example, Figure \ref{FigExample}.

\begin{figure}
[htb]
\begin{center}
\epsfig{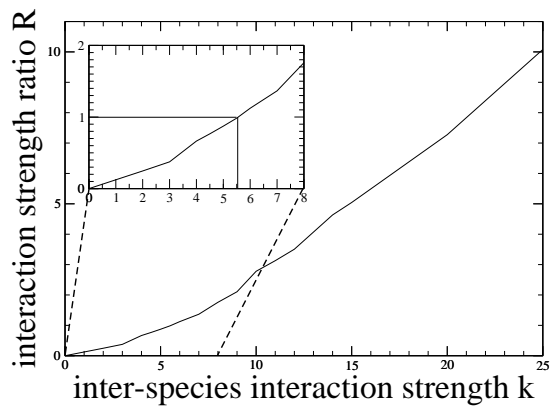}
\caption{\scriptsize{Average value of the ratio of the relative interaction strengths $R$ as a function of $k$, approximately a straight line for $k \ge 8$.} \label{FigKvsKrat}}
\end{center}
\end{figure}

The appearance of diversity is clear when considering the ratio of co-interaction strength to self-interaction strength $R = <\frac{k\sum{J_{ij}n_j}}{\epsilon E_i n_i}>$ shown in Figure \ref{FigKvsKrat}. At $k \approx 5.55$, $R = 1$, meaning that the co-interaction strength is greater than the self interaction strength for $k > 5.55$.  For $k \ge 10$, $R \sim k$ as co-interaction becomes the dominant driving force and selection acts to maintain positive interactions.  Above this $k$, the individual terms in $\sum{J_{ij}n_j}$ and $\epsilon E_i n_i$ are maximised by selection and therefore become independent of $k$ (again ignoring fluctuations).  For $k \le 5.55$, interactions do not contribute to fitness of the wildtype.  For $k \in (5.55,10)$, the relative importance of the two driving forces changes.  Below $k=5.55$ diverse q-ESS states are not found\footnote{The apparent non-unity diversity below the threshold value is due to occasional mutant fluctuations above the wildtype threshold chosen.} as co-interaction is always weaker than self-interaction leading to monodominance.  
\newline 

As soon as it is possible to survive via co-interactions, the selection pressure on self-interaction becomes weaker.  Although it is still beneficial to have the highest self-interaction possible, the additional variable of co-operation pulls species away from the maximally self-interacting peak.

An interesting feature of Figure \ref{FigExample} is that the number of q-ESS switches is high.  At lower interaction strengths, monodominant q-ESSs tend to remain for the entire run, with a low possibility of a switch to another monodominant q-ESS with better self-interaction properties.  As interaction strength increases, the number of q-ESS switches also increases leading to a greater rate of exploration in genotype space.

\subsection{Explanation for a cross-over in diversity}
\label{Jtheory}

We can show the existence off a threshold $k$ value by a simple argument from the definition of $H$.  Consider a system containing only one species, $a$, with genotype label $\mathbf{S}^a$, and an individual $\alpha$ from that species.  Then for all individuals, $H_a^- = H_\alpha^- = \epsilon E_a - \mu n_a$, as there is no other species present to interact with.  We now consider adding an individual $\beta$ of species $b$, thus $N = (n_a + n_b) \approx n_a \gg n_b$, and so:

\begin{align}
H^+_a &= \frac{k n_b J_{a b}}{(n_a + n_b)} + \epsilon E_a \frac{n_a}{(n_a + n_b)} - \mu (n_a + n_b) \\
&\approx \epsilon E_a - \mu n_a 
\end{align}

Similarly for $\beta$ from species $b$:

\begin{align}
H^+_b &= \frac{k n_a J_{b a}}{(n_a + n_b)} + \epsilon E_b \frac{n_b}{(n_a + n_b)} - \mu (n_a + n_b) \\
&\approx k J_{b a} - \mu n_a
\end{align}

For species $b$ to be able to invade the system, we require that $p_{off}^\beta > p_{off}^\alpha$, or equivalently $H^+_b > H^+_a$.  This can be rewritten as:

\begin{equation}
\label{EqnA}
k > \frac{\epsilon E_a}{J_{b a}}
\end{equation}

To prevent $b$ taking over the system and forming a new monodominant state, we also require that in the limit of large $n_b$, species $a$ does not die out completely.  Thus the reverse formula must also be true:

\begin{equation}
\label{EqnB}
k > \frac{\epsilon E_b}{J_{a b}}
\end{equation}

$E_a$ has been selected to be high initially, as $a$ was successful on its own; similarly $J_{b a}$ will be selected to be high to ensure $b$ can proliferate.  Thus the threshold $k$ (called $k_{min}$) is positive (and from Equation \ref{EqnA}, $k_{min} \approx \epsilon$, as both $E_a$ and $J_{b a}$ will be close to $1$).  The more likely restraint on $k_{min}$ comes from Equation \ref{EqnB}, as neither $E_b$ nor $J_{a b}$ have been selected for - we take mean values to get an estimate.  $E_b$ is uniform on $(0,1)$ and thus has mean $0.5$, but $J_{a b}$ must be positive for a steady state to exist, and because of its distribution \cite{TaNaBasic}, has a mean value\footnote{$J$ has a mean of zero, but here we are taking the mean of the positive part of the distribution, which has non-zero mean.} of around $J_{a b} = 0.2$.  We therefore find that the minimum value of $k$ for a diverse state to exist is around $k_{min} = 2.5 \epsilon$, so $k_{min}=5$ in the situation studied in this paper ($\epsilon = 2$).

This means there is a threshold for $k$ below which no diversity can be found, and then increasing $k$ allows for increasing diversity.  We do not expect accuracy in the numbers here, as we have taken no account of fluctuations around the mean values, or stochasticity in the reproduction process.  Still, the match to the observed values is good.

\section{Discussion}

In our model, we have found that there will be, in general, a cross-over to a state with coexistence of diverse types as the interaction increases.  There is a critical value of interaction strength beyond which the monodominance is broken down.  In future work, it will be important to understand the nature of the cross-over theoretically, beyond the naive estimate of the mean-field type calculation given in \ref{Jtheory}.

Each evolutionary course can be different in the simulation. If the initial species has neighbours in genotype space that interact favourably with each other and negatively against the wildtype, then it will quickly go extinct and a number of q-ESS switches are observed.  Other initial conditions allow the interactions of local mutants to favour the wildtype, and monodominance continues for a longer time, possibly beyond the timescale of the simulation.  On a cross-over from one q-ESS state to another, our model predicts that at low interaction strengths, only cross-overs between monodominant states can occur.  However, if the interaction strength $k \ge \epsilon$ then all initial wildtypes should be able to diversify eventually, possibly through a route of alternate monodominant states.  If $k \gg \epsilon$, the contribution to the weight function from the self-interaction becomes negligible and the system reduces to the original Tangled Nature model with the weight function $H_0$, meaning all states are diverse.  In the experiment of \emph{E.coli} \cite{DiverseEcoli}, switches to distinct populations of different types were observed through the course of evolution, which may correspond to q-ESS in the model.  By repeating experiments, this diversification was reproducible, irrespectively of initial conditions, even though the evolution course itself differs by each experiment.  These behaviours are similar to those observed in the simulation.

We have shown theoretically that diversification can occur non-spatially when interactions of sufficient strength are introduced.  The interaction strength $k$ can be interpreted as a density in bacterial systems, as the total flow of chemicals between cells and the media increases, which leads to strong interaction among cells.  Indeed, by increasing the population density, changes to a state with diverse types was observed \cite{DiverseEcoli}.

Our theoretical results enable us to probe the underlying factor allowing diversification that is unobservable in real systems.  Essentially we require both: 

\begin{enumerate}
\item The mean current co-interaction strength is equal to, or greater than, the self-interaction strength; i.e. if a new individual of a new species would gain from interactions with the average population more than another of the same species, then it will flourish, with coexistence even if it is not a better competitor on its own. The greater the ratio of the co-interaction to the self interaction, the less selection pressure acts via the species characteristics (the $\epsilon$ term in Eq. \ref{Hdef}).

\item The possible mutations from the wildtype reinforce themselves, or other species, more than the wildtype.

\end{enumerate}

In the experiment by Kashiwagi et al., a main source of interaction is glutamine.  The glutamine synthesised by `efficient' types of bacteria is leaked out to the medium, through which they interact with each other.  When they experimentally cut off the interaction by adding glutaminase in the medium to eliminate the leaked glutamine, only the fittest remains.  Although this experiment clearly demonstrates the importance of the interaction, the degree of the interaction cannot be determined from the experiment.  Our model result suggests that there is a critical value of interaction beyond which the coexistence appears.  Although the necessity of dense population to show the coexistence in the experiment suggests such kind of critical interaction value, it will be important to obtain it explicitly in the experiment and to study the nature of the cross-over.  Also, according to the theory, the degree of interaction for the coexistence must be of the same order as the internal process, as characterised by the ratio $R$ (see Figure \ref{FigKvsKrat}).  The requirement that $k \approx 3\epsilon$ is due to the values of the mean of (positive couplings) $<J> \approx 0.2$ and $<\epsilon> = 0.5$.  This suggests, for example, that the leaked glutamine from the "efficient" bacteria is of the same order of the amount of that synthesised within.  In fact, in experiments \cite{PrivCom} with higher population of the bacteria with very low enzyme activity, high leakage of glutamine is observed.

Phenotypic plasticity will assist in the second condition if multiple expressions of the wildtype are possible.  Then a mutant may gain by specialising into one of these expressions \cite{KanekoPlasticity}.  This plasticity is present in our model only in that an individual's fitness is a function of the frequency of active genotypes, which are linked in a complex network \cite{TaNaNetwork} created by the interactions.  Thus, plasticity may explain why the random interactions used in the TaNa model provide a good explanation as it may increase differences between separate genotypes.  Indeed, $k$ can be thought of in terms of phenotypic plasticity, as the interactions emphasise the differences between species, with all similarities being absorbed into the $\mu N$ term.  Thus, the existence of $k_{min}$ implies that there is a similar threshold in the plasticity, below which only one expression of a species will be observed but above which, multiple expressions of the same species coexisting may be found.

There is nothing in the model that refers explicitly to bacterial systems, and the effect described here should be observable in macroscopic, sexually reproducing species, assisted by phenotypic plasticity \cite{KanekoPlasticity} and mating preference \cite{EvEcol}. Of course, directly measuring the interaction strength is more difficult in these systems, and the timescales involved makes macroscopic evolutionary experiments difficult.  Models like this should assist in determining which aspects of microbial experiments can be extrapolated to macroscopic evolution.

We have shown that at low interaction strengths, the evolutionary stable state is simply the most efficient competitor dominating. However, at higher interaction strengths, competition for resource is mediated by interspecial interaction and coexistence may occur between varying ability competitors.  Evolution may be faster in this regime as interaction opens up new survival niches, which increase with the diversity and hence increase the available niches again. We have also been able to qualitatively predict the existence of the two regimes from the dynamics in the model.

\section*{\small{Acknowledgements}}

\small{We thank Tetsuya Yomo for discussion on the experiment of \cite{DiverseEcoli} and the Engineering and Physical Sciences Research Council (EPSRC) for funding both Daniel Lawson's PhD, and Kunihiko Kaneko's visit to the U.K.  We also thank Andy Thomas and Gunnar Pruessner for help processing the model and for setting up the BSD cluster, speeding computations enormously.}

\let\oldbibliography\thebibliography
\renewcommand{\thebibliography}[1]{%
  \oldbibliography{#1}%
  \setlength{\itemsep}{0pt}%
}

\begin{spacing}{0}
\bibliographystyle{unsrt}
\begin{scriptsize}
\bibliography{evolution}

\begin{thebibliography}{10}

\bibitem{Biology}
N.~A. Cambell.
\newblock {\em Biology}.
\newblock The Benjamin/Cummings Publishing Company, 2725 Sand Hill Road, Menlo
  Park, California 94025, 1996.

\bibitem{DiverseEcoli}
Akiko Kashiwagi, Wataru Noumachi, Masato Katsuno, Mohammad~T. Alam, Itaru
  Urabe, and Tetsuya Yomo.
\newblock Plasticity of fitness and diversification process during an
  experimental molecular evolution.
\newblock {\em J. Molec. Evol.}, 52:502--509, 2001.

\bibitem{PrivCom}
Kashiwagi et~al.
\newblock Private communication.
\newblock 2005.

\bibitem{TaNaBasic}
Kim Christensen, Simone~A. di~Collobiano, Matt Hall, and Henrik~J. Jensen.
\newblock Tangled nature: A model of evolutionary ecology.
\newblock {\em J. Theor. Biol.}, 216:73--84, 2002.

\bibitem{TaNaQuasi}
Simone~Avogadro di~Collobiano, Kim Christensen, and Henrik~Jeldtoft Jensen.
\newblock The tangled nature model as an evolving quasi-species model.
\newblock {\em J. Phys A}, 36:883--891, 2003.

\bibitem{TaNaTimeDep}
Matt Hall, Kim Christensen, Simone~A. di~Collobiano, and Henrik~Jeldtoft
  Jensen.
\newblock Time-dependent extinction rate and species abundance in a
  tangled-nature model of biological evolution.
\newblock {\em Phys. Rev. E}, 66, 2002.

\bibitem{NeutralTheory}
Stephen Hubbel.
\newblock {\em The Unified Neutral Theory of Biodiversity and Biogeography}.
\newblock Princeton University Press, 41 William Street, Princeton, New Jersey
  08540, 2001.

\bibitem{NTecologyNature}
Igor Volkov, Jayanth~R. Banavar, Stephen~P. Hubbell, and Amos Marita.
\newblock Neutral theory and relative species abundance in ecology.
\newblock {\em Nature}, 424:1035--1037, 2003.

\bibitem{McKaneReview}
Ricard~V. Sol\'{e}, David Alonso, and Alan McKane.
\newblock Self-organised instability in complex ecosystems.
\newblock {\em Phil. Trans. R. Soc. Lond. B}, 357:667--681, 2002.

\bibitem{SpecDiv}
Michael~L. Rosenzweig.
\newblock {\em Species diversity in space and time}.
\newblock Cambridge University Press, The Edinburgh Building, Cambridge CB2
  2RU, 1995.

\bibitem{KanekoPlasticity}
Kunihiko Kaneko.
\newblock Symiotic sympatric speciation: consequence of interaction-driven
  phenotype differentiation through developmental plasticity.
\newblock {\em Popul. Ecol.}, 44:71=85, 2002.

\bibitem{TaNaNetwork}
Paul Anderson and Henrik~Jeldtoft Jensen.
\newblock Network properties, species abundance and evolution in a model of
  evolutionary ecology.
\newblock {\em Journal of Theoretical Biology}, 232:551--558, 2005.

\bibitem{EvEcol}
Erik~R. Pianka.
\newblock {\em Evolutionary Ecology}.
\newblock Addison Wesley Educational Publishers, 1301 Sansome St, San
  Francisco, CA 94111, 2000.

\end{thebibliography}
\end{scriptsize}
\end{spacing}
\end{document}